\documentclass[twocolumn,showpacs,preprintnumbers,amsmath,amssymb,floatfix]{revtex4}
\usepackage{graphicx}% Include figure files
\usepackage{dcolumn}% Align table columns on decimal point
\usepackage{bm}% bold math

\newcommand\be{\begin{equation}}
\newcommand\ee{\end{equation}}
\newcommand\bea{\begin{eqnarray}}
\newcommand\eea{\end{eqnarray}}
\newcommand\ba{\begin{array}}
\newcommand\ea{\end{array}}
\def\simlt{\stackrel{<}{{}_\sim}}
\def\simgt{\stackrel{>}{{}_\sim}}
\begin{document}

\preprint{IFT-UAM/CSIC-07-01, UAB-FT/623}

\title{Novel Effects in Electroweak Breaking from a Hidden
Sector}

\author{Jos\'e Ram\'on Espinosa$^{1}$ and Mariano Quir\'os$^{2}$ }
\affiliation{$$}
\affiliation{
$^{1}${IFT-UAM/CSIC, Cantoblanco, 28049 Madrid, SPAIN}\\
$^{2}${ICREA/IFAE, UAB 08193-Bellaterra Barcelona, SPAIN}}

\date{\today}% It is always \today, today,

\begin{abstract}
The Higgs boson offers a unique window to hidden sector fields $S_i$,
singlets under the Standard Model gauge group, via the renormalizable
interactions $|H|^2 S_i^2$. We prove that such interactions can provide
new patterns for electroweak breaking, including radiative breaking by
dimensional transmutation consistent with LEP bounds, and trigger the
strong enough first order phase transition required by electroweak
baryogenesis.
\end{abstract}

\maketitle

{\bf 1. Introduction.}
The Standard Model (SM) of electroweak
and strong interactions can not be considered as a fundamental theory,
since it fails to provide an answer to many open questions (the
hierarchy, cosmological constant and flavor problems, the origin of
baryons, the Dark Matter and Dark Energy of the Universe, \dots), but
rather as an effective theory with a physical cutoff $\Lambda$ that most
likely shall be probed at the LHC experiment. Many SM extensions,
e.g.~string theory, contain hidden sectors with a matter content
transforming non-trivially under a hidden sector gauge group but
singlet under the SM gauge group. It has recently been noticed that
the SM Higgs field $H$ plays a very special role with respect to such
hidden sector since it can provide a window (a portal~\cite{hidden0})
into it through the renormalizable interaction $|H|^2 S_i^2$ where
the bosons $S_i$ are SM singlets.

This coupling to the hidden sector can have important implications both
theoretically and for LHC phenomenology as has been discussed in recent
literature~\cite{hidden0,hidden1,hidden2,hidden3,hidden4,hidden5,hidden6,hiding}.  In this letter we show that the presence of a
hidden sector may have dramatic consequences for electroweak symmetry
breaking (in particular it enables new patterns of electroweak
symmetry breaking, including radiative breaking by dimensional
transmutation consistent with present LEP bounds on the Higgs mass)
and for electroweak baryogenesis (it makes easy to get a 
first order phase transition as strong as required for electroweak
baryogenesis). Furthermore, under mild assumptions those hidden
sector fields are stable and can constitute the Dark Matter of the
Universe.

{\bf 2. Electroweak breaking.}  We will consider a set of $N$
fields $S_i$ coupled to the SM Higgs doublet by the
(tree-level) potential
\be V_0=\,m^2H^\dagger H+\lambda\,(H^\dagger H)^2+\zeta^2 H^\dagger H
\sum_i S_i^2 .
\label{0pot}
\ee
We will assume for the moment that the fields $S_i$ are massless so
they only will get a mass from electroweak breaking.  In the
background Higgs field configuration defined by $\langle
H^0\rangle=h/\sqrt{2}$, the one-loop effective potential (in Landau
gauge and $\overline{MS}$ scheme) is given by
\be
V=\frac{m^2}{2} h^2 + \frac{\lambda}{4} h^4
+\sum_\alpha \frac{N_\alpha M_\alpha^4}{64\pi^2}
\left[\ln\frac{M_\alpha^2}{Q^2}-C_\alpha\right]\ ,
\label{1pot}
\ee
where $\alpha=\{S,Z,W,t,h,G\}$ for singlet hidden sector fields, gauge
bosons, top, Higgs and Goldstones respectively, with
$N_\alpha=\{N,3,6,-12,1,3\}$. Inspired by the case of stops, we
choose $N=12$ for our numerical work. Next, $C_\alpha=3/2$ for
fermions or scalars and $5/6$ for gauge bosons, and the $h$-dependent
masses are $M_S^2=\zeta^2h^2$, $M_Z^2=(g^2+g'^2)h^2/4$,
$M_W^2=g^2h^2/4$, $M_t^2=h_t^2h^2/2$, $M_h^2=3\lambda h^2 + m^2$,
$M_G^2=\lambda h^2 + m^2$. The renormalization scale $Q$ enters
explicitly in the one-loop logarithmic correction and implicitly
through the dependence of all couplings and fields on $t=\ln Q$ in
such a way that $dV/dt=0$ is satisfied. For now we simply choose the
scale as $Q=M_t(v)$ and fix the parameters (at that scale) to get $\langle
h\rangle=v\simeq 246$ GeV.

For $\zeta^2 < h_t^2/2\simeq 0.65$ the one-loop term in (\ref{1pot}) is 
dominated by 
the standard top contribution but for $\zeta^2 > h_t^2/2$ hidden scalars 
start to dominate. The structure of the effective potential is best 
described by using
Fig.~\ref{phasespace}. Consider first the $(\zeta,\lambda)$-plane in
the upper plot. Besides the lines of constant $M_h$, we can
distinguish four regions. {\bf i)} The region below the blue line
[defined by $V''(v)=0$] is forbidden: there $M_h^2<0$.
The extremal at $h=v$ is a maximum that degenerates into an
inflection point on the blue line.  {\bf ii)} In the region above
the blue line but below the red line there is an electroweak minimum, but 
it is a
false minimum with respect to the (true) minimum at the origin. The
red line is defined by $V(v)=V(0)$, {\it i.e.}  both minima, at the
origin and at $h=v$, are degenerate on that line. This region ii) is
therefore unphysical without a mechanism to populate the metastable
minimum (in general, the true minimum at the origin would be preferred
at high temperature and the electroweak transition would never take
place).  {\bf iii)} In the region above the red line but below the 
green line [defined by $V''(0)=0$] the electroweak minimum is stable and 
there is
a barrier separating the false minimum at the origin from the
electroweak minimum ($m^2>0$). This region is
very interesting for two reasons:

\begin{figure}
\includegraphics[width=8cm,height=5.5cm]{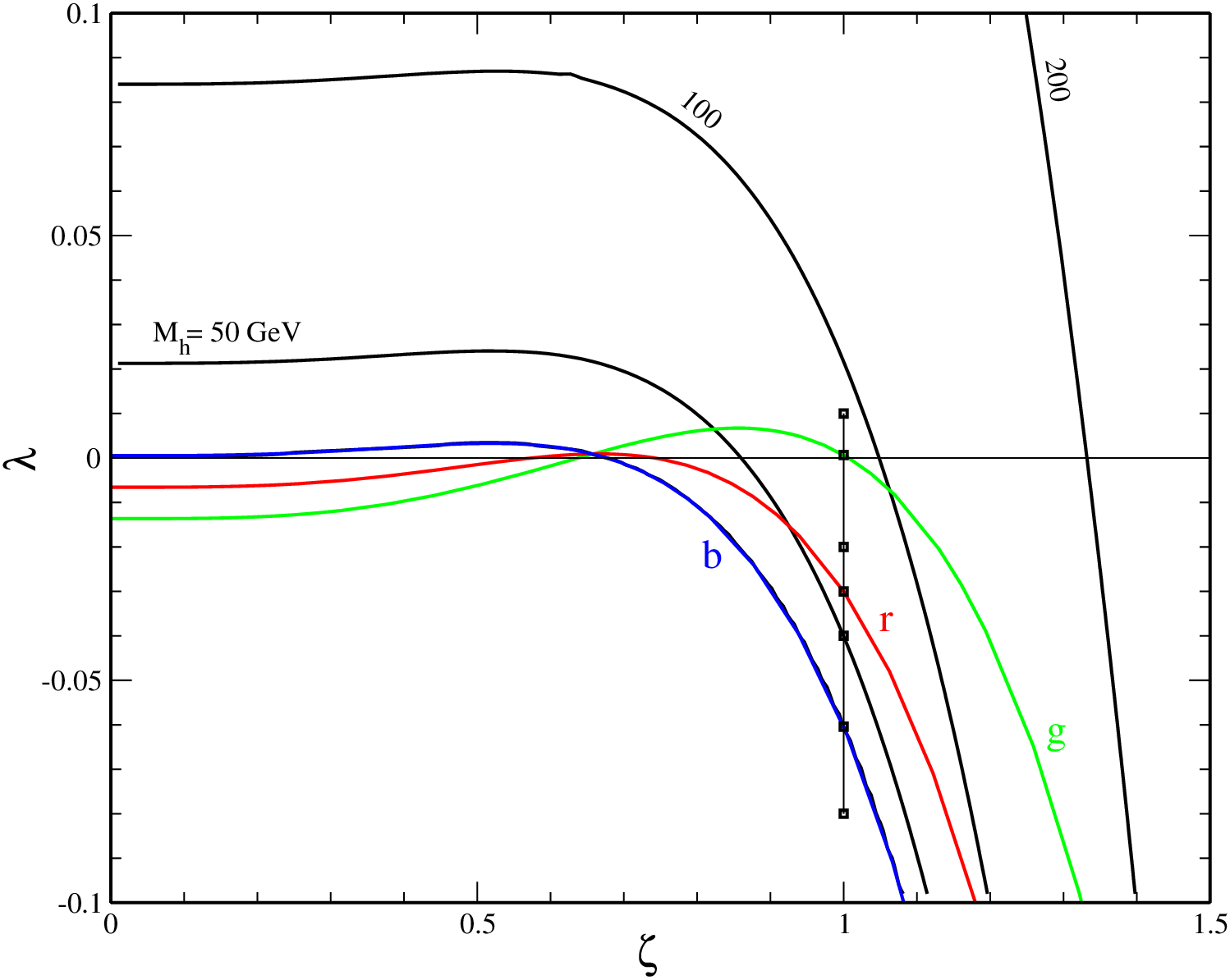}\\[0.5cm]
\includegraphics[width=8cm,height=5.5cm]{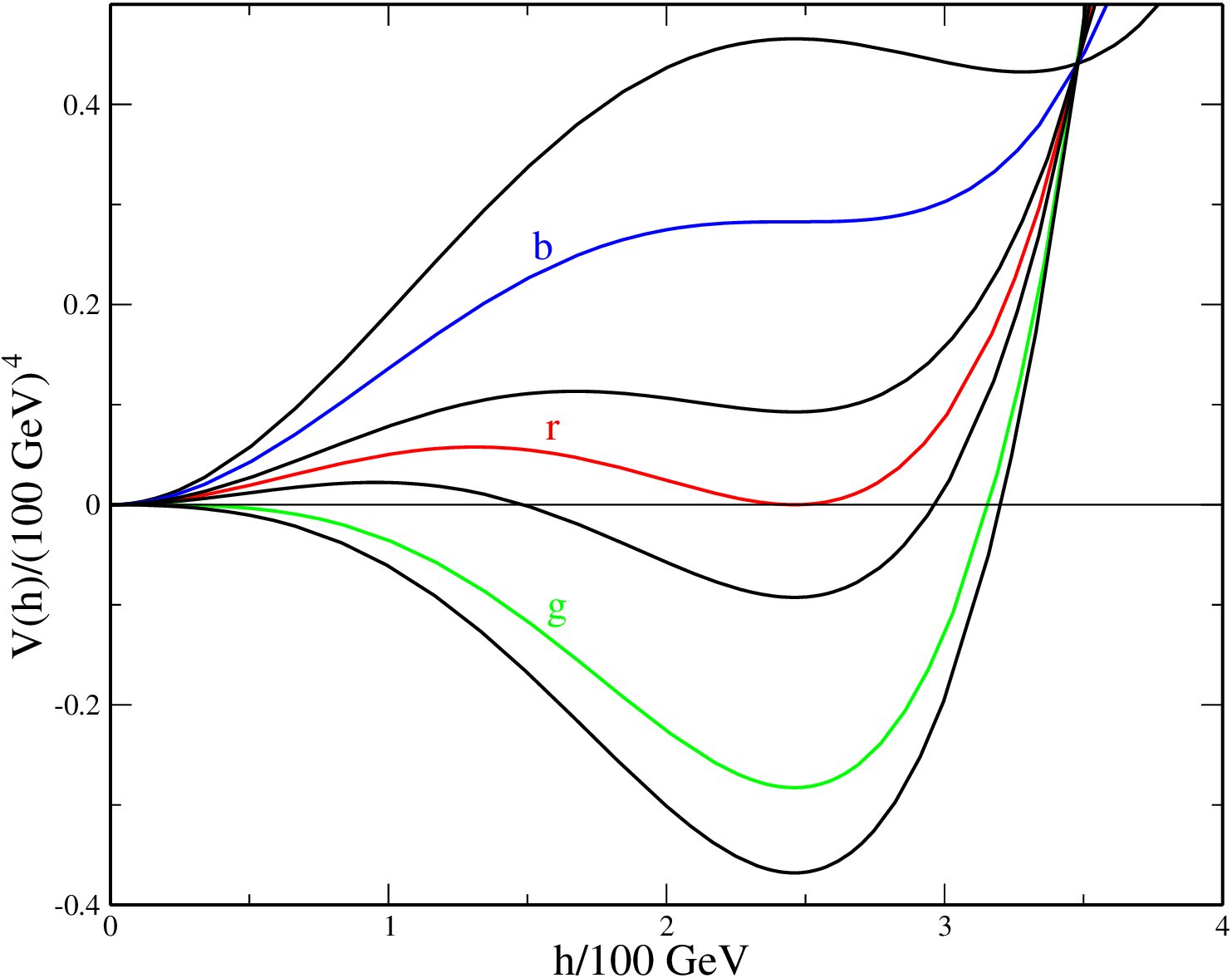}
\caption{\label{phasespace}  Upper plot: In the plane $(\zeta,\lambda)$,
the green line corresponds to the condition $V''(0)=0$, the red
to $V(v)=V(0)$ and the blue to $V''(v)=0$. Black solid lines
correspond to the indicated values of $M_h$.  Lower plot: Potential
for $\zeta=1.0$ and different values of $\lambda$ (or $M_h$) as marked on 
the vertical line in upper plot.}
\end{figure}

\begin{itemize}
\item The barrier between both minima (at zero temperature) will
  produce an overcooling of the Higgs field at the origin at finite
  temperature, strengthening the first order phase transition (see
  below).
\item 
  Electroweak symmetry breaking is not associated with the
  presence of a tachyonic mass at the origin, as in the SM. Instead
  it is triggered by radiative corrections via the mechanism of 
  dimensional transmutation.
\end{itemize}
The minimum at the origin becomes a maximum at the green line. In fact the 
green line corresponds to the conformal case where
$m^2=0$ and electroweak breaking proceeds by pure dimensional
transmutation (see also \cite{Nicolai}).  {\bf iv)} Finally, in the region
above the green line the origin is a maximum as in the SM, with $m^2<0$.

Notice that, while $\lambda>0$ is required in the SM case ($\zeta=0$ 
axis), now $\lambda<0$ is accessible for sufficiently large
$\zeta$. The shape of the potential for the different cases is
illustrated by the lower plot of Fig.~\ref{phasespace}, where
$\zeta=1$ has been fixed and we vary $\lambda$ as indicated by the
vertical line in the upper plot of Fig.~\ref{phasespace}. From
bottom-up the potentials have decreasing values of $\lambda$. The
lowest potential corresponds to $\lambda=0.01$ and has the
conventional maximum at the origin. The green potential corresponds to
the conformal case where $m^2=0$ (in this particular example also 
$\lambda$ is zero!). The next line corresponds to $\lambda=-0.02$ with a 
barrier
between the origin and the electroweak minimum while for the red
potential the two minima become degenerate. The next line corresponds
to the potential for $\lambda=-0.04$ where the electroweak minimum is
already a false minimum, which becomes an inflection point at the blue
line where $M_h=0$. Finally the highest line corresponds to
$\lambda=-0.08$ and the electroweak extremal is a maximum (the
potential has a minimum somewhere else, for some $\langle h\rangle >
v$. If $\zeta^2$ were smaller, $\zeta^2\simlt h_t^2/2$,
the potential would instead be destabilized due to $\lambda<0$.). 

\begin{figure}
\includegraphics[width=8cm,height=5.5cm]{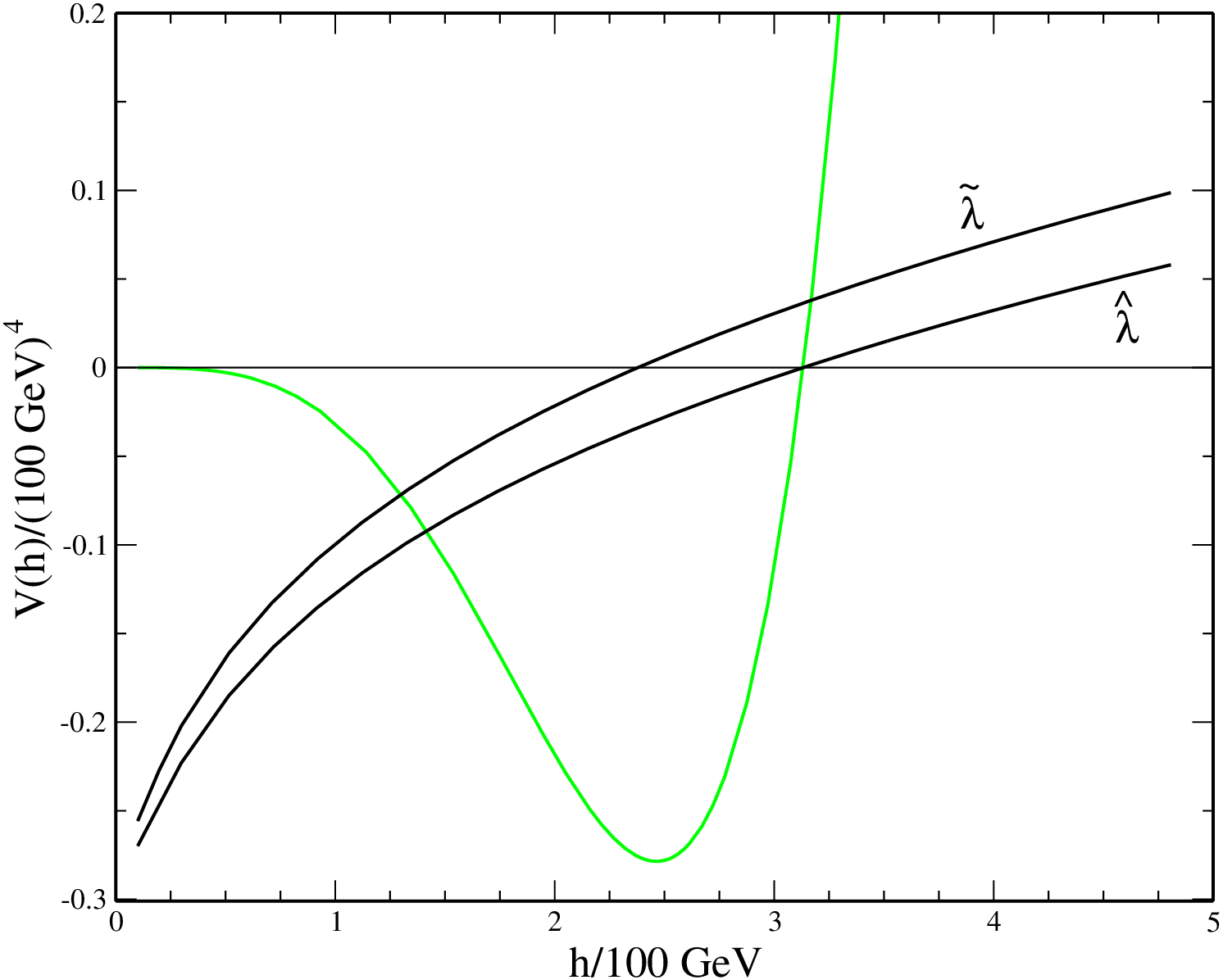}
\caption{\label{running}  
Green: Effective potential for the conformal case.
Black: running $\tilde\lambda$ and $\hat\lambda$, with $Q=M_t(h)$.}
\end{figure}

In order to have a better understanding of the phenomenon of radiative
electroweak breaking by dimensional transmutation in this setting
consider the conformal case with $m^2=0$. Then improve the one-loop
effective potential of Eq.~(\ref{1pot}) by including the running with
the renormalization scale of couplings and wave functions. We use for
that the SM renormalization group equations (RGEs) supplemented by the
effects of $S_i$ loops plus the RGEs for the new couplings to the
hidden sector (see~\cite{eqPRD} for details). The RGE-improved
effective potential is scale independent and we can take advantage of
that to take $Q=M_t(h)$ as a convenient choice to evaluate the
potential at the field value $h$ (with all couplings ran to that
particular renormalization scale). This results in a ``tree-level''
approximation $V\simeq (1/4)\hat\lambda h^4$ with \cite{lambdaeff}
\be
\hat\lambda\equiv \lambda + \sum_\alpha
\frac{N_\alpha \kappa_\alpha^2}{64\pi^2}
\left[\ln\frac{\kappa_\alpha}{h_t^2}-C_\alpha\right]\ ,
\label{lambdahat}
\ee
where the $\kappa_\alpha$'s are coupling constants, defined by the
masses as $M_\alpha^2=(1/2)\kappa_\alpha h^2$. The behavior of the
one-loop potential as a function of $h$ is captured by the
``tree-level'' approximation above through the running of
$\hat\lambda$ with the renormalization scale, linked to a running with
$h$ by the choice $Q=M_t(h)$. To illustrate this, we show in
Fig.~\ref{running} the effective potential for this
conformal case (green lines in Fig.~\ref{phasespace}) with $m^2=0$ and
$\zeta=1$, together with the effective quartic coupling
$\hat\lambda(h)$. We can see that the scale of dimensional
transmutation is related to the scale at which the potential
crosses through zero. The structure of the potential is then
determined by the evolution of $\hat\lambda$: for small $h$,
$\hat\lambda<0$ destabilizes the origin while, for larger $h$,
$\hat\lambda>0$ stabilizes the potential curving it upwards in the
usual way.

We can define a different effective coupling, $\tilde\lambda$, by the
approximation $\partial V/\partial h\simeq \tilde\lambda h^3$, which fixes
$\tilde\lambda$ to be given by (\ref{lambdahat}) with $C_\alpha\rightarrow
C_\alpha-1/2$. Fig.~\ref{running} shows that $\tilde\lambda$ crosses
through zero precisely at the minimum of the potential. This shows then
how the electroweak scale is generated by dimensional transmutation: a
suitably defined effective quartic Higgs coupling turns from positive to
negative values, with $v$ given by the implicit condition
$\tilde\lambda(v)=0$. Needless to say, such running of $\tilde\lambda$ 
would not be possible in the SM and is due to the effect of $\zeta$ in the 
RGEs, which counterbalances the effect of $h_t$.

{\bf 3. Electroweak phase transition.} In the presence of hidden
sector fields $S_i$ coupled to the SM Higgs as in Eq.~(\ref{0pot}) the
electroweak phase transition is strengthened by: {\it a)} The 
thermal contribution from $S_i$, if $\zeta$ is large enough. This 
fact was known already~\cite{Anderson:1991zb,Espinosa:1993bs}. {\it b)} The
fact that, in part of the $(\zeta,\lambda)$-plane, there is a barrier
separating the origin (energetically favored at high temperature) and
the electroweak minimum at zero temperature. This effect is
new~\cite{barrier}.

To study the strength of the phase transition we consider the
effective potential at finite temperature, $T$.  In the one-loop
approximation and after resumming hard-thermal loops for Matsubara
zero modes, the thermal correction to the effective potential $\Delta
V_T$ is given by
\bea
\frac{T^4}{2\pi^2}\sum_{\alpha}N_\alpha\int_0^\infty
dx\ x^2\log\left[1-\varepsilon_\alpha
  e^{-\sqrt{x^2+M_\alpha^2/T^2}}\right]&&
\label{Tpot}\nonumber\\
+\frac{T}{12\pi}\sum_{\alpha}\frac{1+
\varepsilon_\alpha}{2}N_\alpha \left\{M^3_\alpha -
  \left[ M^2_\alpha+\Pi_\alpha(T^2)\right]^{3/2}  \right\}\ 
,&& \label{Tbos}
\eea
where $\varepsilon_\alpha=+1(-1)$ for bosons (fermions) and
$\Pi_\alpha(T^2)$ is the thermal mass of the corresponding field (for
more details see Ref.~\cite{eqPRD}). The considered approximation is
good enough for our purposes since, as we will see, the phase
transition is strongly first order and mainly driven by the
contribution to the thermal potential of the $S_i$ fields for which
the thermal screening $\Pi_S$ is enough to solve the infrared
problem. Notice that the second term in Eq.~(\ref{Tbos}), responsible
for the thermal barrier, takes care of the thermal resummation for 
bosonic zero modes.

\begin{figure}
\includegraphics[width=8cm,height=5.5cm]{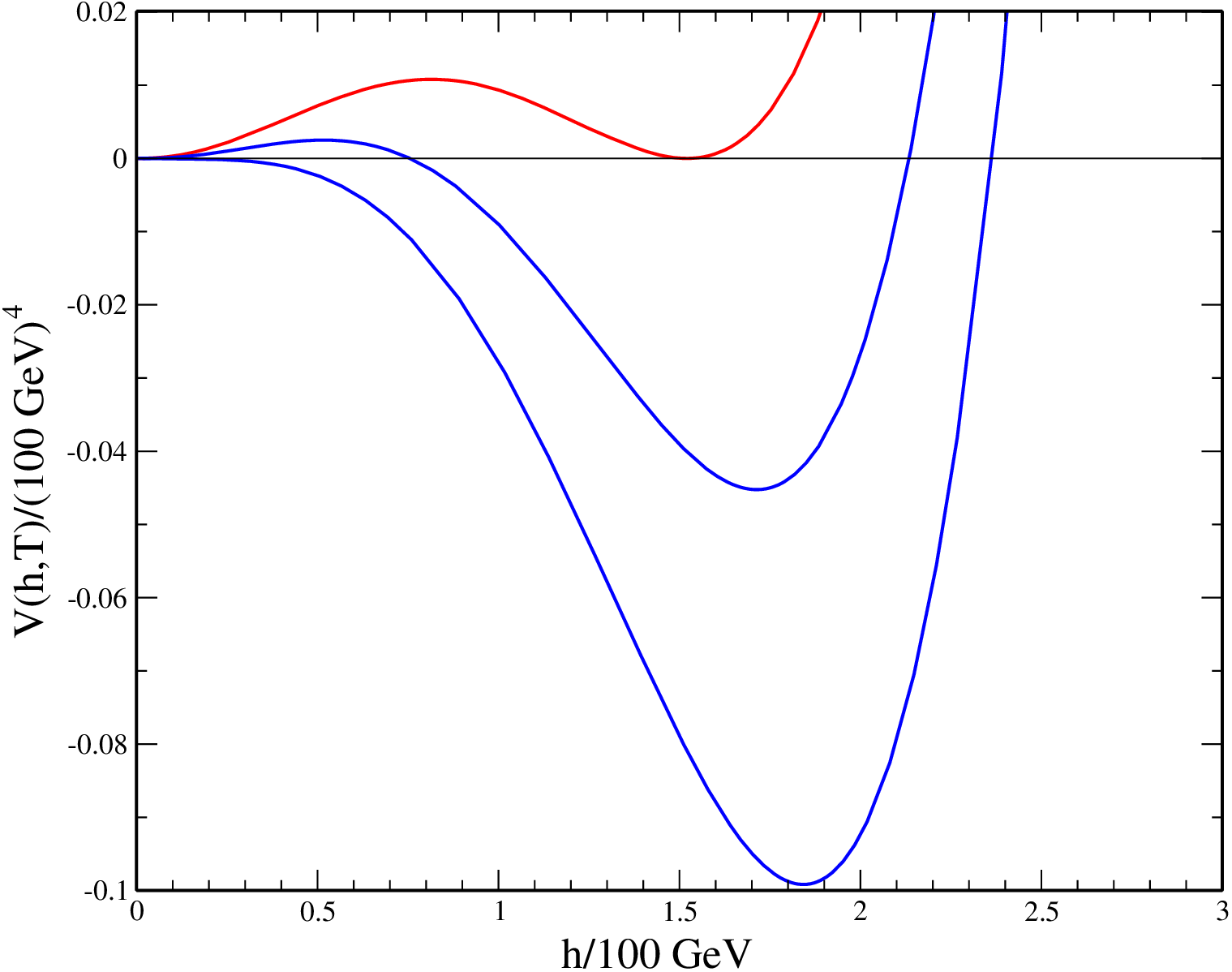}\\[0.5cm]
\includegraphics[width=8cm,height=5.5cm]{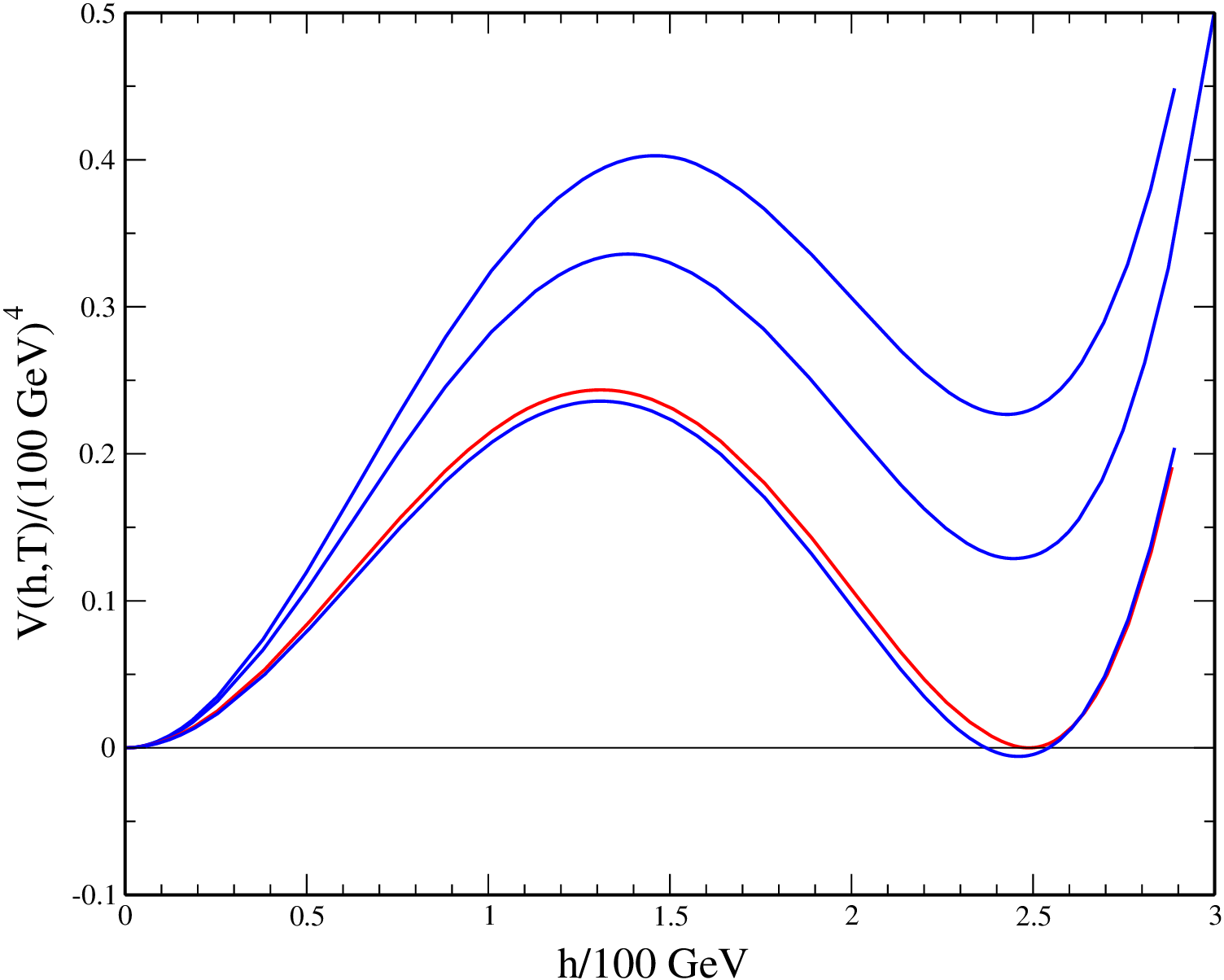}
\caption{\label{phase}
Effective potential around the EW phase transition, for $M_h=125$ GeV.
Upper plot: $\zeta=0.8$ and $T=110.85$, 108.00 and
105.00 GeV, with $R\simeq 1.37$.
Lower plot: Same for $\zeta=1.365$ and $T=50.00$, 40.00, 30.08
and 0 GeV with $R\simeq 8$. }
\end{figure}

We define $T_c$ as the critical temperature at
which the origin and the non-trivial minimum at $\langle
h(T_c)\rangle$ become degenerate, calling its ratio $R\equiv\langle
h(T_c)\rangle/T_c$. The baryogenesis condition for non-erasure of the
previously generated baryon asymmetry requires $R\simgt
1$~\cite{Bochkarev:1990gb}. In general, identifying the critical
temperature with the real tunneling temperature (which is smaller)
underestimates $R$ so that our approximation provides a conservative
estimate of the order parameter $R$. For a more detailed analysis see
Ref.~\cite{eqPRD}.

\begin{figure}
\includegraphics[width=8cm,height=5.5cm]{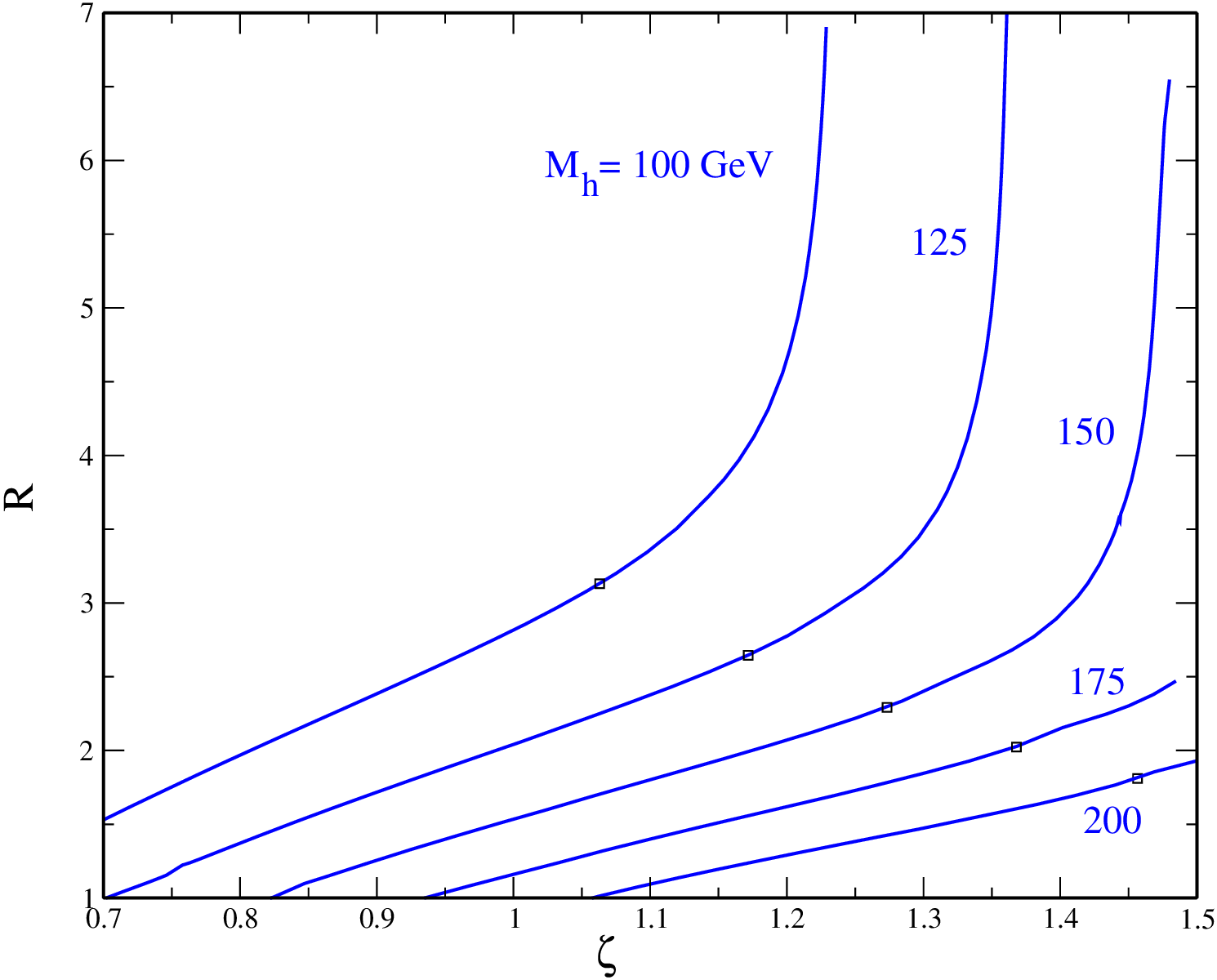}
\caption{\label{R}
$R\equiv \langle h(T_c)\rangle/T_c$ as a function of $\zeta$ for several
values of $M_h$, as indicated.}
\end{figure}

We illustrate in Fig.~\ref{phase} the behavior of the effective potential
around the critical temperature for a fixed Higgs mass ($M_h=125$ GeV) and
for two typical cases. In the upper plot we consider a case where the
strength of the phase transition is only due to the thermal barrier from
$S_i$ fields (with $\zeta=0.8$) with no $T=0$ barrier, leading to 
$R\simeq 1.37$. In the lower plot,
with $\zeta=1.365$, the barrier persists all the way down to $T=0$ making
the value of $R$ much larger ($R\simeq 8$).  The dependence of $R$ with 
$\zeta$ for
different values of $M_h$ is displayed in Fig.~\ref{R} where the strong
enhancement in the values of $R$ produced inside the region where the
barrier between the origin and the electroweak minima persists at $T=0$ is
apparent (the square dots mark in each case the region beyond which there
is a barrier at $T=0$).  The answer to the general question of what is the
upper bound on the Higgs mass to avoid baryon asymmetry washout depends on
how large $\zeta$ can be, which in turn depends on the cutoff $\Lambda$. A
low cutoff, e.g.~$\Lambda\sim 1-10$ TeV, allows values of $\zeta$ up to
$1.3-1.8$ while a higher cutoff $\Lambda\sim 10^{5}$ GeV would only allow
values of $\zeta\simlt 1$.

A pending issue is how the baryon asymmetry is created (perhaps by the 
hidden sector) since within the SM the amount of CP violation, given by 
the
CKM phase, is admittedly insufficient~\cite{Gavela:1994dt} (although a
way out associated with physics solving the flavor problem at a
high-scale was proposed in~\cite{Berkooz:2004kx}). An interesting
possibility from the low energy point of view is the appearance of
CP-violating effective operators. For instance the dimension-six
operator $g^2 |H|^2 F\tilde F/(32\pi^2\Lambda^2)$ can
generate the baryon-to-entropy ratio (for maximal CP
violation)~\cite{Dine:1990fj}
$n_B/s\sim 3.1\kappa\times 10^{-9}\left(T_c/\Lambda\right)^2$,
where $\kappa\simeq 0.01-1$, which is roughly consistent with WMAP
data for $\Lambda$ in the TeV range.

{\bf 4. Conclusion.} In this letter we have explored new and
dramatic effects that a hidden sector, singlet under the SM gauge
group, can have concerning electroweak symmetry breaking and
electroweak baryogenesis. Completely new patterns for the Higgs
potential and new ways of radiative breaking by dimensional
transmutation are found, some of them indirectly leading to a very
strong EW first order phase transition. For such a strong first-order
phase transition the model can provide a strong signature in
gravitational waves~\cite{Randall:2006py}. Moreover if the hidden
sector has a global $U(1)$ symmetry that guarantees the stability of
$S_i$-scalars (as we are assuming) and some subsector of it has a
large invariant mass it can also provide good candidates for Dark
Matter~\cite{McDonald,eqPRD}.

\begin{acknowledgments}
Work supported in part by CICYT, Spain, under
contracts FPA2004-02015 and FPA2005-02211; by a Comunidad de Madrid
project (P-ESP-00346); and by the European Commission under contracts
MRTN-CT-2004-503369 and MRTN-CT-2006-035863. J.R.E. thanks CERN for
partial financial support during the final stages of this work.
\end{acknowledgments}

\end{document}